\documentclass[twocolumn,aps,prl,showpacs]{revtex4-1}
\usepackage{graphicx,epsf,epsfig}
\usepackage{dcolumn}
\usepackage{bm}
\usepackage{amsmath, amsthm, amssymb}
\usepackage{color}
\usepackage{ulem}
\usepackage{accents}

\usepackage{enumitem}
\setlist[itemize]{noitemsep}
\usepackage{epstopdf}
\definecolor{pink}{rgb}{1,0.5,0.5}
\allowdisplaybreaks
\begin{document}

\title{Transmission engineering as a route to subthermal switching}
\author{Avik W. Ghosh}
\affiliation{Department of Electrical and Computer Engineering \\
University of Virginia, Charlottesville, VA 22904 \\
}%
\date{\today}
\begin{abstract}
The physics of low subthreshold devices is interpreted in terms of a gate dependent change in their mode averaged transmission function, in addition to a capacitive shift in their overall mode spectrum. Accordingly, we explore a variety of subthermal switches that alter the bandwidth, bandgap or amplitude of the transmission, and combinations thereof. The Landauer theory for current flow provides a convenient way to derive the subthreshold swing in each case analytically and suggests ways to beat the Boltzmann limit.
\end{abstract}
\maketitle

\begin{figure}[ht!]
\begin{center}
\includegraphics[width=3.3in,height=3.3in]{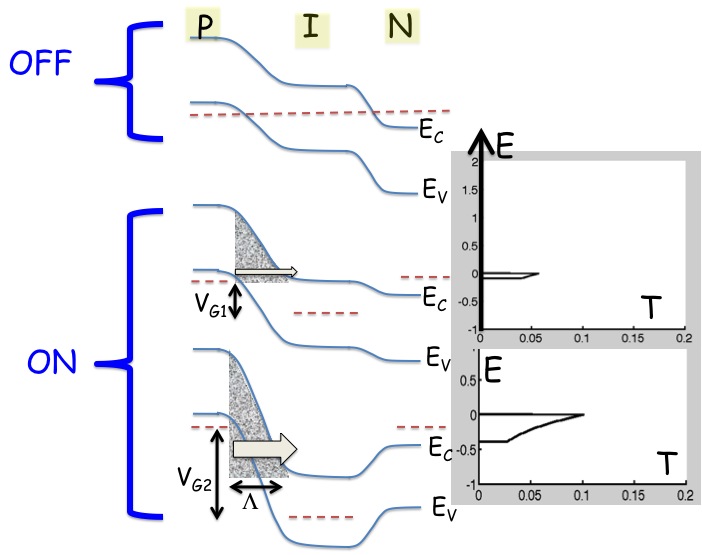}
\end{center}
\caption{Top to bottom: Schematic band diagrams of PIN junction under OFF, slightly above threshold at $V_{G1}$ and fully ON at $V_{G2} > V_{G1}$. Red dashed lines show local quasiFermi levels. As the intrinsic conduction band edge $E^I_C$ sweeps past the valence band edge $E^P_V$ of the P region, the transmission window increases in bandwidth and amplitude with progressively increasing band to band tunneling through a rapidly narrowing triangular barrier, shown in grey. \textcolor{black}{The simulations to the right are based on a WKB approximation through a barrier with gate split 10 nm, bandgap 1.1eV, dopings $10^{16}cm^{-3}$ and $10^{12}cm^{-3}$ across the P-I region, tunneling effective mass of 0.19$m_0$.
The subthreshold swing $S \approx S_0\Delta\Phi/k_BT$ where $\Delta\Phi = E_V^P-E_C^I$.}} 
\label{f1}
\end{figure}

\begin{figure}
\begin{center}
\includegraphics[width=3.3in,height=3.1in]{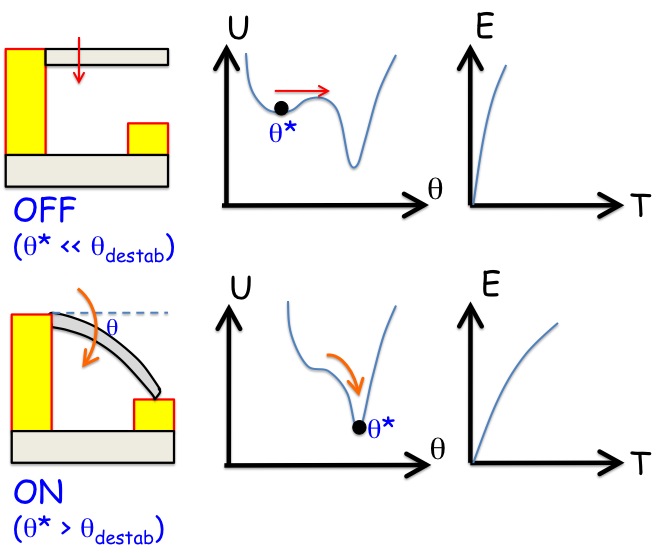}
\includegraphics[width=2.4in,height=2.0in]{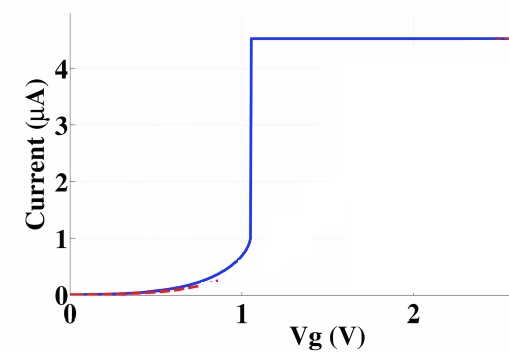}
\end{center}
\caption{A nanomechanical relay turns on through a phase transition when the gate bias shifts its elastic potential minimum into a second minimum given by electrostatics (capacitive plus Van Der Waals sticking). The elimination of the intermediate barrier leads to an abrupt switch from metastable to stable state as the relay bends over and sticks, exponentially increasing the tunnel transmission. The gate dependence of the tunnel transmission leads to a very sharp subthreshold swing $S \approx S_0(U_0^{Bend}/k_BT)q({\theta^*-\theta_{\rm{destab}}})/\mu\kappa$, where $\mu$ is the dipole moment and $\kappa$ is the tunnel decay constant (we are assuming here the dipolar drive is stronger than the capacitive drive). Just around destabilization, there is a finite temperature smearing set by the correlation between the destabilization force and the transmission function. \textcolor{black}{Simulated steep transfer characteristic at bottom (Fig. reproduced from Unluer, Ghosh, unpublished).}}
\label{f4}
\end{figure}

\begin{figure}
\begin{center}
\includegraphics[width=3.3in,height=2.8in]{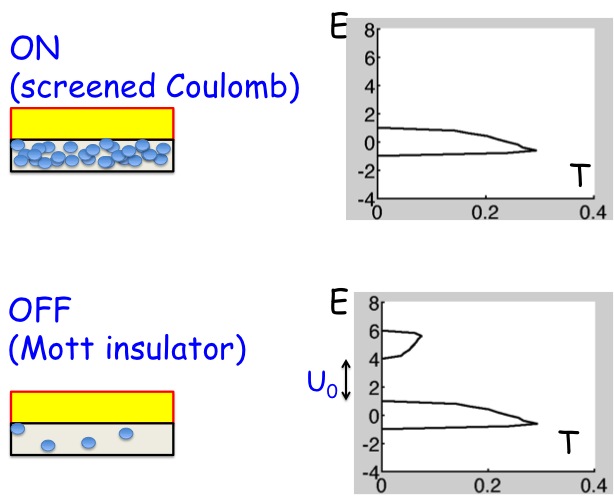}
\end{center}
\caption{A metal enjoys a screened Coulomb interaction. Depleting charges enhances their Coulomb repulsion until the bare Coulomb energy $U_0$ exceeds the level broadening, whereupon an abrupt Mott transition occurs to an insulating state. 
{Simulations are shown, with each self-consistent Hubbard band with a 2eV bandwidth included in a Newns-Anderson model \cite{na1},~\cite{na2} and a single electron charging energy toggling between a fully screened value of 0~eV and an unscreened value of 5~eV.}}
\label{f5}
\end{figure}

{\subsection{1. The high cost of switching}}
As we approach the end of the roadmap for complementary metal oxide semiconductor (CMOS) devices, a critical limiter for 
continued scaling is the power dissipation in a transistor. 
The act of charging a gate capacitor $C$ with a
power supply voltage of $V_D$ leads to a dissipated energy of $\int_0^{CV_D} VdQ = CV_D^2/2$ over the full charging period. This dissipation arises because the power supply delivers at a constant voltage $V_D$ while the voltage stored on the gate ramps from zero 
to $V_D$. Assuming a sharp turn on and off compared to the RC response time of the device, the balance of energy
is irretrievably lost as Joule heat in the wires.  
Summed over a complete charging and discharging cycle, the corresponding power dissipation becomes $\alpha NCV_D^2f$
where $f$ is 
the clock frequency, $N$ is the number of gates and $\alpha < 1$ is the activity factor for square wave switches. Including typical numbers from the semiconductor roadmap, we find that an end of the roadmap device operating under conventional Dennard scaling would dissipate at least $100$ MW/cm$^2$ of power beyond 2020, which far exceeds the known limits on heat removal capability. 
\\\\
\textcolor{black}{The dissipated energy can be written as the product of charge times voltage, $Q\Delta V$. While the number of charges $Q$ has reduced over generations of Dennard scaling, the voltage has been limited by thermal fluctuations to $\sim k_BT/q$ and is hard to scale.}
Industry has already addressed the dissipation problem by abandoning Dennard scaling, in particular the scaling of clock frequency (even though the cut-off frequency keeps increasing), and using multicore strategies to restore the computational power. As a result Moore's Law, cast in broader terms of functions per dollar, is still preserved. Much of the progress has been in controlling short channel electrostatics (e.g. high k oxides, trigates, FinFETs), reducing the footprint, and improving mobility (e.g. Ge or III-V channels). However, much more innovative architectural and software solutions are needed to control the rapidly increasing energy budget. It is thus contextual and potentially useful to explore novel physical mechanisms for switching that can lower the energy cost for the same functionality.  Developing a switch that operates at very low voltage $V_D$ would go a long way towards alleviating this problem. 
\\\\
The voltage needed for binary switching, i.e., moving a charge reliably from source to drain, is determined by two factors, (i) the number of decades that the current must swing through in the subthreshold regime (i.e., the ON/OFF ratio), and (ii) the steepness of the gate transfer characteristic that converts that ON/OFF ratio to a corresponding voltage swing. The ON/OFF ratio is set by the imposed standards for circuit level error rates (and independently by the need for a high ON current to reduce delay and low OFF current to reduce standby dissipation). The subthreshold swing on the other hand is set by fundamental physics, specifically the rate of spontaneous thermal excitation over an energy barrier, related ultimately to the acceptable write error rate. 
\\\\The focus of this paper is on the physics of the subthreshold swing and ways to beat the Landauer limit. \textcolor{black}{This alone may not suffice in building a winning device, as the corresponding power gain must not come at the expense of speed, i.e., current drive (more precisely we need to minimize the action, which amounts to minimizing the free-energy times delay product)}. Ultimate scaled devices are often swamped by contact effects and parasitic capacitances, which can easily mask an intrinsically superior switching curve. Fabrication issues notwithstanding, we believe however that understanding the physics of subthermal switching has inherent intellectual merit, and duly optimized may even have practical relevance.
\\\\
{\subsection{2. The Landauer equation as a starting point}}
Conventional switches operate by raising and depressing the potential on a gate electrostatically, thereby allowing
charges to jump over the barrier through thermal emission. For a well defined energy barrier, the rate of current generation with gate bias is set by the Boltzmann  tail of the  Fermi-Dirac distribution of electrons in the contacts, ultimately
translating to a subthreshold swing of $k_BT\ln{10}/q \approx 60$ mV/decade (the logarithm arises from converting
Boltzmann exponentials into decades). It therefore stands to reason that one way to bypass this  Shannon-Landauer limit \textcolor{black}{\cite{shannon},~\cite{shannon2}} and squeeze out more current than usual under the same operating voltage, would be to make the transmitting barrier itself respond nontrivially 
to the applied gate bias. 
\\\\The Landauer equation \cite{datta,agbook,agbook2} gives us a way to capture the barrier dynamics quantitatively. The conductance
is set by the total quantum mechanical transmission over the propagating modes within an energy window set by the applied drain bias.
\begin{equation}
\textcolor{black}{I = \frac{q}{h}\int_{-\infty}^{\infty} dE {T}(E,V_G)M(E)\Bigl(f_1(E)- f_2(E)\Bigr)}
\end{equation}
where $M(E)$ is the mode count in the device including spin, and $f(E)$ is the equilibrium Fermi-Dirac distribution with subindices $1,2$ referring to the bias separated  electrochemical potentials  $\mu_{1,2}$ in the source and drain contacts. ${T}(E,V_G)$ is the mode averaged quantum mechanical transmission function. 
\\\\The Landauer equation is quite powerful in that it smoothly interpolates between conductance quantization in a purely ballistic device ${T} = 1$, tunneling in a thin barrier ${T} \sim \exp{(-2\kappa L)}$ and Ohmic conduction in a dirty diffusive sample ${T} = \lambda/L$. The different limits can be derived from a {\it{single}} resonant transmission formula  across two scatterers in series. \textcolor{black}{The sum over all possible histories of the electron transiting between two  scatterers, going through multiple reflections and transmissions, leads to a geometric series. As a result, the coherently summed transmission has the form ${T} \sim 1/(a + b\cos\theta)$}. The decay constant $\kappa$, \textcolor{black}{phase dependent scattering angle $\theta$}, and scattering length $\lambda$ are in general energy dependent, and often gate voltage dependent, adding more complexity to the underlying physics. 
\\\\
In a regular transistor, the role of the gate is to electrostatically shift the mode spectrum $M$ relative to the contact Fermi energies, in other words, alter their relative band alignment along the energy axis. To quantify,
it helps to rewrite the Landauer equation in an exactly equivalent but lesser known form \cite{1bagwell}, which starts with the {\it{zero temperature}} current and then reintroduces the Fermi tail as a convolution with a thermal broadening function $F_T(E) = -\partial f(E)/\partial E$, 
\begin{equation}
I = \frac{q}{h}\int_{\mu_1}^{\mu_2} dE {T}\Bigl[M\otimes F_T\Bigr]
\end{equation}
The equation tells us rightaway what limits switching. Since the convolution of two functions is limited by the degree of spread of the slower function, the most abrupt turn on we can hope to achieve is if the mode spectrum were to change as a step function. In that ideal case, the abruptness is limited by the gate dependent shift of the thermal broadening function relative to the mode spectrum. In the subthreshold regime, $f \approx e^{-\beta E}$, $\beta = 1/k_BT$, so that $F_T \approx \beta f$. Assuming the best case scenario where the modes simply shift rigidly with a capacitive gate transfer factor $\alpha_G$ in addition to an abrupt turn on, $M(E,V_G) = M_0\Theta(E-q\alpha_GV_G)$ \textcolor{black}{where $\Theta$ is the step function},  we get the normalized transconductance, equal to the inverse subthreshold swing 
\begin{equation}
S^{-1} = g_m = \displaystyle\frac{\partial \log_{10}{I}}{\partial V_G}=  q\alpha_G\beta/\ln{10}
\end{equation}
leading to the well recognized expression for the subthreshold swing, $S = S_0 = k_BT\ln{10}/q\alpha_G$, i.e., 60 mV/decade for perfect gate control $\alpha_G = 1$. It is worth emphasizing that even this ideal limit needs special conditions, namely \textcolor{black}{a sharp turn on for the mode spectrum, and near ideal electrostatics where the gate holds the channel charge constant}. As long as the conduction band is filled by electrons following $f$ and valence band by holes following $1-f$, we are stuck with this best case limit. 
\\\\
In the above derivation, we have ignored a possible gate voltage dependence of the transmission function ${T}(E,V_G)$ itself, which can add another term
\begin{equation}
S^{-1} =  (q\alpha_G\beta + \overline{\partial\ln{T}/\partial V_G})/\ln{10}
\end{equation}
where the bar denotes that the second term is mode averaged over the energy bias window \textcolor{black}{(It is important to look at the modal sum to see the overall trade-off)}. We see that the subthreshold swing is reduced if the \textcolor{black}{section of the transmission energy curve bounded by the Fermi window increases in area under the action of a gate voltage}. This would happen for example if the system were to shrink its bandgap or open additional conducting channels under a gate bias, in addition to the band realignment. 
\\\\
In the following sections, we look at various ways to engineer such a gate voltage dependent transmission function. 
\\\\
{\subsection{3. Gating the transmission bandwidth and amplitude: Tunnel Field Effect Transistors (TFETs)}} 
Let us start with a homojunction TFET, constructed by doping a material into a p-i-n regime. The gate bias is used to pull the conduction band edge of the intrinsic region past the valence band edge of the p type region to enable band to band tunneling (Fig.~\ref{f1}). We now have an anomalous region where the conduction band of the intrinsic region can be populated by the electrons from the valence band of the p region. As a result, the net electron population in the intrinsic region becomes non thermal with a sharp energy cut off created by the band filtering action of the two, effectively cooling the electrons and reducing their subthreshold swing. Assuming a triangular potential in the WKB approximation for tunneling, the transmission modulated by the added band filter function \cite{2knoch}
\begin{equation}
T_{WKB}  \approx e^{\displaystyle -4\sqrt{2m}U_0^{3/2}/3\hbar{\cal{E}}_0}
\Biggl[\Theta(E -E_V^P) -\Theta(E -E^I_C)\Biggr]
\label{twkb}
\end{equation}
with barrier height $U_0 = E_G + \Delta\Phi$, $E^I_C = E_V^P - \Delta\Phi$,  and electric
field ${\cal{E}}_0 = U_0/q\Lambda$. The current is obtained by invoking Landauer equation, summing the WKB transmission over the transverse modes using 2-D Fermi functions, leading to the Simmons equation for tunneling \cite{simmons}.
\\\\
Note that the Landauer current now has two window functions, one set by the gate and drain voltage dependent
Fermi functions, and the other by the relative locations of the conduction and valence bands. The
actual window is thus decided by the overlap of those two window functions, and this overlap
changes abruptly since the band edges themselves do not come with any thermal smearing (in practice,
they would smear due to the presence of phonons and traps). From this current we can then calculate the subthreshold swing. As discussed in Knoch {\it{et al}} \cite{2knoch}, there will be two contributions to this, one that involves controlling the transmission window, and the other involves modulating the amplitude of the mode averaged transmission because the barrier becomes more triangular and thinner as a result 
(Fig.~\ref{f1})
\begin{eqnarray}
S^{-1} &=& \displaystyle \frac{\partial \log_{10}{I}}{\partial \Delta \Phi} \approx \Biggl(\underbrace{{\Bigl[f(E-E_V^P) - f(E-E_C^I)\Bigr]}}_{\displaystyle\text{Bandwidth~Modulation}}\nonumber\\
&+&
\underbrace{4\sqrt{2m}q\Lambda E_G^{3/2}/3\hbar{{U}}^2_0}_{\displaystyle\text{Amplitude~Modulation}}\Biggr)\frac{1}{\ln{10}}
\end{eqnarray}
The first term in the bracket gives us $S \sim \Delta \Phi\ln{10}/q$. As the two band
edges sweep past each other, $\Delta\Phi$ and thus the subthreshold swing rises from zero and reaches a finite value. 
\\\\
In practice, the subthreshold swing is limited by phonon assisted tunneling that can initiate even within the bandgaps \cite{st3,koswatta}. Besides, the challenge is to maintain a low subthreshold swing over a large enough voltage swing, while also maintaining a large tunnel ON current. One way to get the latter is to have different bandgaps on both sides using a staggered heterostructure \cite{st1,st2}. This becomes a problem of material band engineering, especially since heterojunctions tend to form interface traps that can alter the band alignment in a nontrivial way and provide additional leakage paths. 
\\\\
{\subsection{4. Gating transmission gap: 2D PN junctions}} 
To get a change in current, what we need ultimately is a transmission gap and not an outright bandgap.
In other words, we need momentum rather than energy filtering. For TFETs, the problem is that the ON
state is still limited by tunneling. In contrast, graphene and other Dirac cone systems (such as Bi2Se3 surfaces)
have zero bandgap, which is great for the ON current but not for the OFF.  A zero gap however has the advantage that opening a gap somehow (e.g. electrostatically) leads to a large fractional change in conductance. Thus, opening a transmission gap
electrostatically not only reduces the OFF current, but in addition the gate tunability of this transmission gap would invariably generate 
a low subthreshold swing. As the Fermi energy moves towards a bandedge, that gap closes in addition,
pulling out more current for the same gate voltage. 
\\\\
A transmission gap based on momentum filtering can be engineered using electrostatically gated PN junctions in 2D chiral systems. In these materials the bandstructure resembles photons and follows a Dirac cone. Furthermore, there is a strong locking between electron momentum and the chiral degree of freedom \cite{rkastnelson,rredwanprb,rti}, which for graphene is a pseudospin (the phase angle describing the mixing of the pz dimer orbitals) and for 3D topological insulators is the actual spin. Because of the conservation of transverse momenta and pseudospin across the PN junction, the transmitted modes lie within a cone whose angle depends on the voltage barrier across the junction. \textcolor{black}{In fact, the electron trajectories follow Snell's law, with the local gate induced potential playing the role of refractive index}. The Bloch wavefunction for a graphene electron in 
a minimal dimer p$_z$ orbital basis set is a 2 $\times$ 1 pseudospinor $(1, \pm e^{\displaystyle i\theta})$, where $\theta = \tan^{-1}(k_y/k_x)$ is the angle of the electron trajectory (+ for conduction band, - for valence band). Matching the
wavefunction as well as the transverse momentum vector across the junction, we get an angle dependent transmission as \cite{3redwanapl}
\begin{equation}
T_0(\theta_I) = \displaystyle\frac{2\cos{\theta_I}\cos{\theta_T}}{1+\cos{(\theta_I + \theta_T)}}e^{\displaystyle - 2\pi dk_{FI}k_{FT}\sin{\theta_I}\sin{\theta_T}/(k_{FI}+k_{FT})}
\label{Tpnj}
\end{equation}
where `I' and `T' refer to the incident and transmitted modes, related by Snell's law, and $d$ is the separation between two backgates used to electrostatically dope the two segments of the PN junction individually. The current is obtained by integrating over the incident angle modes. 
\\\\
Note that the transmission is explicitly gate voltage dependent, with the barrier $q\Delta V_G = \hbar v_F(|k_{FI}|+|k_{FT}|)$. There is a critical angle $\theta_C \approx 1/\sqrt{\pi k_{F\parallel}d}$
where $\parallel$ refers to the parallel combination of momenta in the exponent in Eq.~\ref{Tpnj}.
However, the gate voltage dependence at this stage is still modest.
\begin{figure}
\begin{center}
\includegraphics[width=3.8in,height=3.3in]{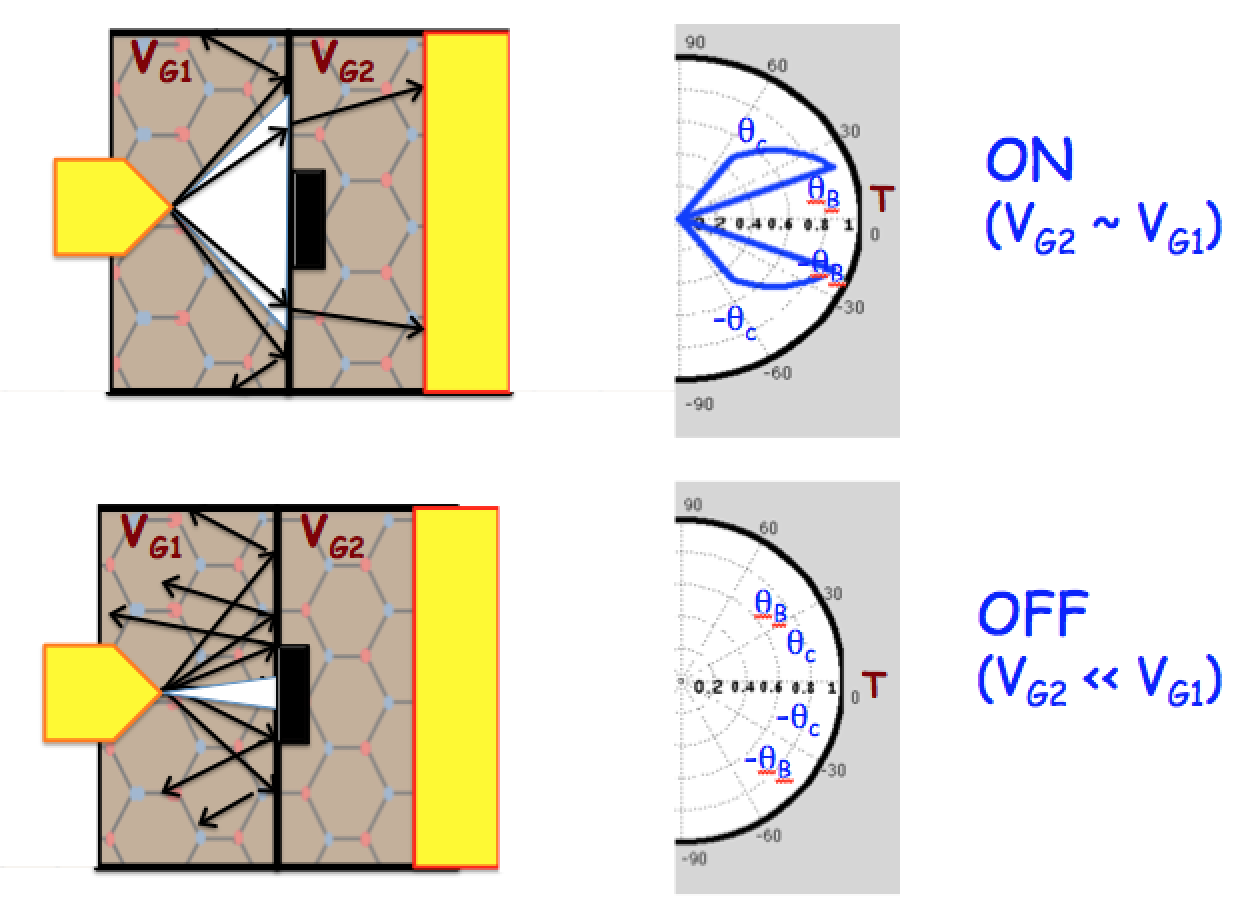}
\includegraphics[width=2.3in,height=2.3in]{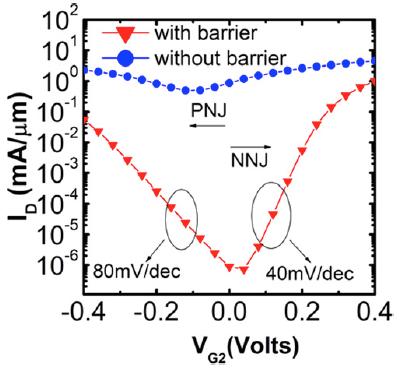}
\end{center}
\caption{Split gated electrostatically doped graphene PN junction (GPNJ) allows electrons to flow through a narrow cone (white triangle) whose critical angle $\theta_C \sim 1/\sqrt{\pi k_{F\parallel}d}$ is set by the gate voltage difference across the junction. However, part of this triangle gets blocked by a separate physical barrier such as a cut (black rectangle) which subtends an angle $\theta_B$ at a point source. For near homogeneous doping $V_{G2} \sim V_{G1}$, the critical angle $\theta_C \sim \pi/2$ exceeds the angle $\theta_B$ and the device is ON. For strongly heterogeneous doping $V_{G2} \ll V_{G1}$, the critical angle falls below the barrier angle $\theta_B$ and all modes are eliminated either by total reflection or by the barrier (device is OFF). The transmission plotted against incident angle  shows a gate tunable bandwidth. Replotted along the total energy axis (not shown) we get a gate tunable transmission gap for all energies where $\theta_C(V_{G2}) < \theta_B$. The result is a low subthreshold swing $S = S_0(1-\sin\theta_B)$ as we ramp the gate voltage in either region from intrinsic towards homogeneous. \textcolor{black}{The bottom panel shows the drastic improvement in simulated transfer characteristics in graphene by using a PN junction and a cut (Fig. reproduced from \cite{3redwanapl}).}}
\label{f2}
\end{figure}
\begin{figure}
\begin{center}
\includegraphics[width=3.3in,height=3.1in]{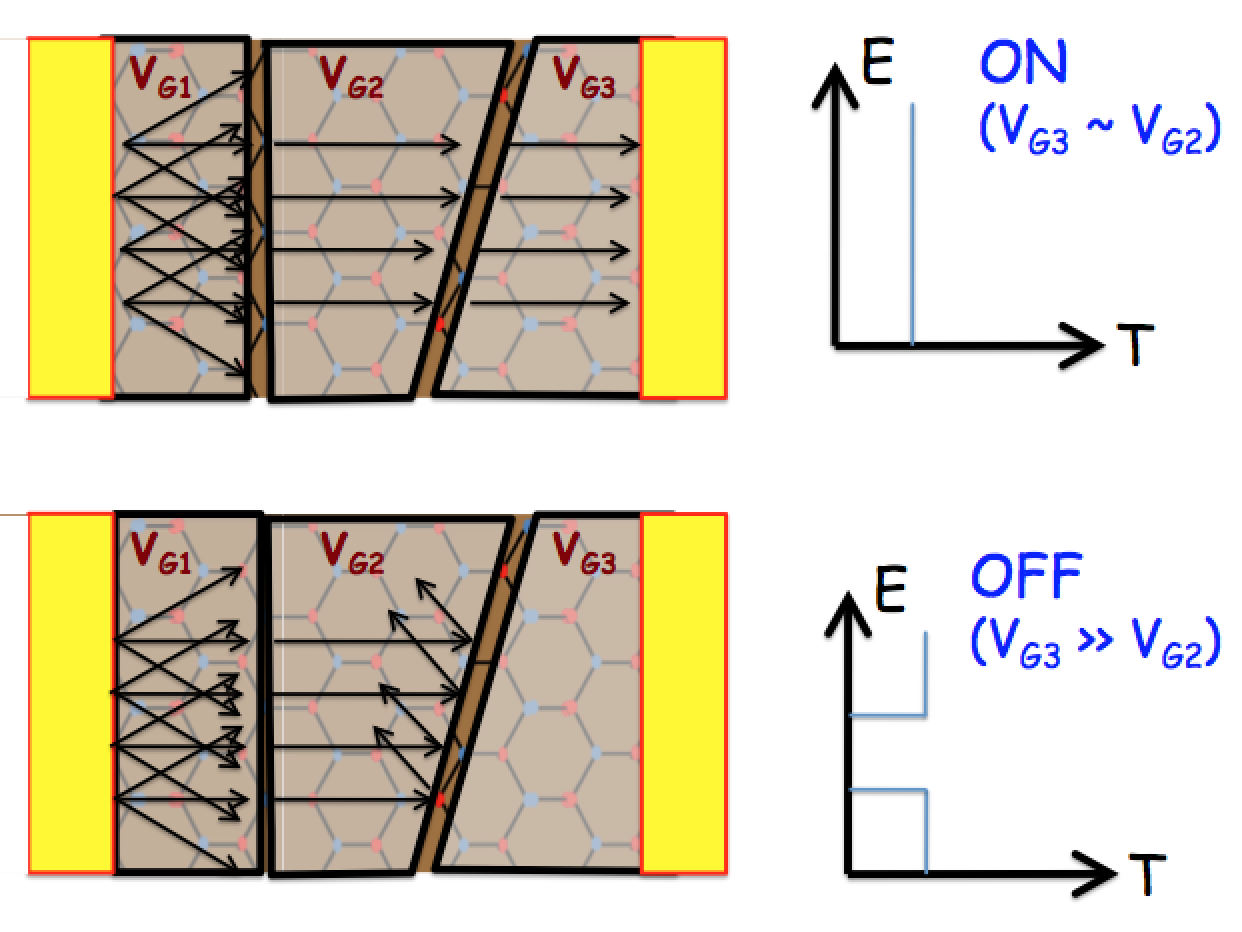}
\includegraphics[width=3.3in,height=3.0in]{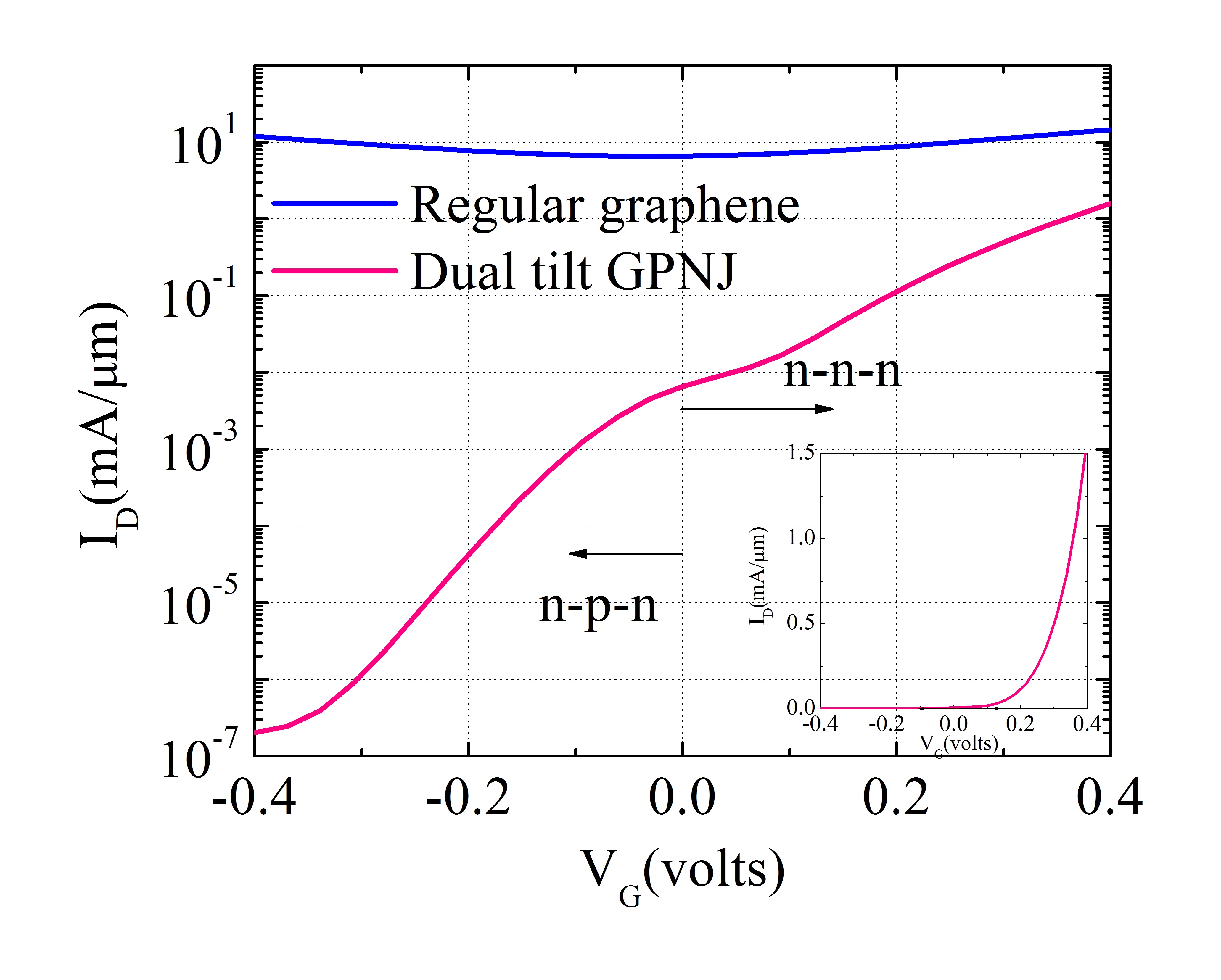}
\end{center}
\caption{Momentum collimation by a sequence of split-gated GPNJs. For homogeneous gating, the electrons go through. By ramping up the second junction voltage, $V_{G3} >> V_{G2}$, the modes are reflected back as the second junction only transmits collimated modes perpendicular to itself. A transmission gap opens whenever the critical angle for the highest mode $\theta_C$ falls below the tilt angle $\delta$, giving a gate tunable transmission gap and a corresponding low subthreshold swing  $S = S_0(1-\sin\delta)$. \textcolor{black}{Simulations are shown at the bottom for a tilted structure, assuming minimal edge reflection and reflectionless contacts (Fig. reproduced from \cite{4redwanacs}).}}
\label{f3}
\end{figure}
 If the maximal mode can now be suppressed geometrically, we will get a complete quenching of the OFF current. One way to do this is to put an added barrier, such as a  cut, inside the graphene sheet at the PN junction (Fig.~\ref{f2}), which subtends an angle $\theta_B$ at an injecting source, assumed to be a collimating point contact.
 If the critical angle $\theta_C$ corresponding to the maximum transmitted mode is smaller than $\theta_B$, i.e. $\theta_C < \theta_B$, then all modes are rejected back to the source and $T$ has an abrupt cut off over a range
of angles and energies
\begin{equation}
T(\theta_I,V_{G1},V_{G2}) = T_0(\theta_I)\Theta(\theta_I-\theta_B)
\label{tc}
\end{equation}
We can also accomplish this quenching with an extended rather than a point contact using a second PN junction tilted at angle $\delta$ relative to the first (Fig.~\ref{f3}). If $\theta_C < \delta$, once again the charges are reflected back to the source, with $\theta_B \rightarrow \delta$ \cite{4redwanacs}.
\\\\ In both cases, we have a range of energies corresponding to $\theta_C < \theta_B, \delta$ where there is no transmission.
This provides a low OFF current while only marginally degrading the ON current. But in addition, the transmission gap is directly controlled by the gate dependent voltage barrier across the junction, $E_G = q\Delta V_G(2\sin\delta/\cos^2\delta)$, vanishing when we reach the homogeneous limit. The gate tunability of the transmission gap leads to a voltage upconversion of the applied gate bias. The conduction and valence band edges shift by different amounts,
$E _{V,C} = \mp q\Delta V_G\sin\delta/(1\pm \sin\delta)$. From this added shift of band edges, we can extract the final subthreshold swing
\begin{equation}
S = k_BT\ln{10}/q \times (1 \pm \sin\delta) 
\end{equation}
with a higher degraded threshold as we move from intrinsic PI to the heterogeneous PN region and a lower improved threshold when we move towards the homogeneous PP region. 
\\\\
A sequence of graphene PN junctions can thus eliminate the undesirable electrons by redirecting them towards
the source. The challenge however is edge scattering, which can mess up momentum filtering by bouncing around the rejected electrons, increasing their chance for being normally incident at the second interface upon repeat attempts. Contact resistances are also problematic with graphene, as are quantum capacitances that tend to stretch out the transfer characteristic, diluting intrinsic improvements. Still, large scale atomic simulations including contact and edge effects suggest that for good contacts with resistances $\sim 100 \Omega-\mu$m, the ON/OFF ratio can be pushed up to at least a couple of orders of magnitude while simultaneously maintaining a high mobility. 
\\\\
{\subsection{5. Gating transmission amplitude: Ion channel relays vs Nano Electromechanical field effect transistors (NEMFETs)}} 
The problem with subthreshold operation is usually the OFF current, which for a CMOS device is now compromised by tunneling and other parasitic transport pathways. One way to ensure a low OFF is to create a mechanical relay that physically moves away from the contacts in order to suppress conduction. The current in these devices is set by electron tunneling into a drain  from the end of a movable cantilever clamped at the source. \\\\The subthreshold swing for relays has been measured to be as low as 0.1mV/decade \cite{5ionescu}, while swings only about 7 mV/decade have been reported for many years in sodium channel switches \cite{6mackinnon}. 
Curiously, the physics in the two cases is different and have distinct origins. The first, operational in NEMFETs, involves a phase transition when the capacitive destabilization energy increases as the channel moves towards the drain and overcomes the elastic restoring forces (Fig.~\ref{f4}). This destabilization can be further aided by short range Van Der Waals pull-in forces. The transition leads to an abrupt change of conformation as the cantilever bends over and sticks to the drain, increasing the current suddenly as a result. The second mechanism, operational in ion channels, relies on several charges moving together for the price of one. In other words, the cantilever functions as a correlated system with a single giant charge, so that the effective single particle destabilization energy decreases rapidly (the associated reduction in power dissipation because of correlated entities is a strong motivation for studying other correlated systems such as magnets and Mott insulators).  Since the tunnel current increases exponential with proximity of the cantilever to the drain, both the NEMFET and the ion-channel mechanisms (phase transition vs correlation) are very promising for low subthreshold behavior. 
\\\\
Ignoring thermal fluctuations, the current for a stiff cantilever is now given by 
\begin{equation}
I = \frac{q}{h}\int_{\mu_1}^{\mu_2} dE {T}\Bigl(E,\theta^*(V_G)\Bigr)M(E,V_G)\otimes F_T(E)
\end{equation}
where the defining angle is obtained from the minimum in the cantilever potential $U(\theta)$, by setting $dU(\theta^*)/d\theta^* = 0$. \textcolor{black}{Let us outline this for a typical cantilever geometry}, with cantilever length $L$, set at a height $h$ above the drain (Fig.~\ref{f4}), setting a maximum angular deformation $\theta_0$ from the horizontal position. The cantilever potential $U$ has contributions from elastic restoration in the form of a bending potential $U_0^{Bend}(\theta -\theta_0)^2/2$, the movable gate capacitance with an air gap $\epsilon A/t_{air}$, dipolar torques $-\vec{\mu}\times\vec{\nabla{V}}$ acting on charges on the cantilever, and Van Der Waals forces enabling adhesion to the drain upon contact. We assume the transmission is dominated by tunneling from the drain to the end of the cantilever, with a decay constant $\kappa$. The normalized transconductance then becomes
\begin{equation}
\bar{g}_m = \displaystyle  \frac{\partial \ln{I}}{\partial V_G}= \beta + 2\kappa L\Biggl(\displaystyle\frac{d\theta^*}{dV_G}\Biggr)
\label{tnems}
\end{equation}
where $L$ is the length of the cantilever. In other words, the electrostatic and conformational parts act in parallel and bring down the overall swing $S = \bar{g}_m^{-1}$ \cite{7ghoshnl},\cite{8ghoshdincer}
i.e.,  
\begin{equation}
S^{-1} \approx S^{-1}_{el} + S^{-1}_{conf}
 \end{equation}
Simplifying the expression with the terms in $U$,  we find that the  conformational part at its minimum
is given by
\begin{equation}
S^{-1}\Bigl|_{Min} = \bar{g}_m\Bigr|_{Max} =  \kappa L \displaystyle\frac{Q_{tot}}{U_0^{Bend}}\frac{\theta_0}{\theta^*-\theta_{\rm{destab}}}
\label{tnems}
\end{equation}
where $Q_{tot} = \partial U/\partial V_G$ is the total charge stored in the cantilever, $\theta_0$ is the full angular deflection to reach the drain, and the destabilization energy $\theta_{destab}$ is given by setting $d^2U/d(\theta^*)^2 = 0$, i.e. the point where the cantilever reaches a point of inflexion between the minimum from the elastic restoring force and that from the electrostatic (capacitive + Van Der Waals + dipolar) destabilization. 
\begin{equation}
(\theta)_{\rm{destab}}= \displaystyle \Biggl[\frac{U_0^{Bend} + 3k_{VDW}}{3(U_0^{Bend} + k_{VDW})}\Biggr]\theta_0
\end{equation}
where $k_{VDW}$ is the nonlinear (angle-dependent) stiffness constant representing the Van Der Waals pull in.
\\\\
The transconductance above diverges near the destablization point from the abrupt phase transition where the electrostatic forces overcome the elastic force, leading to a corresponding abrupt pull in and vanishing subthreshold swing. In practice, the sharpness of the transition will be mitigated by the lack of sharpness when the two wells (the elastic and electrostatic) collide, and there will be a spread $\Delta\theta$ given by the Boltzmann distribution around the point of inflexion. The Landauer current will now need to average over intermediate angles, $T(\theta^*) \rightarrow \int d\theta T(\theta)\exp{[-\beta U(\theta)]}/\int d\theta \exp{[-\beta U(\theta)]}$.
The conformational subthreshold swing can be written as
\begin{equation}\
S_{conf} = \displaystyle\Biggl(\frac{k_BT}{q}\ln{10}\Biggr)\Biggl\langle\frac{qt_{air}(\theta)}{\mu C(\theta)}\Biggr\rangle
\end{equation}
where $\mu$ is the dipole moment and $C$ is a correlation function between the transmission and the gate transfer factor. As we can see from this expression, there are two origins for low subthreshold swing which relate to the NEMFET vs the ion-channel limits. In the former, the additional `pull-in' sucks in the end of the cantilever once it is physically proximal with the drain. The short-ranged Van der Waals force and the longer ranged capacitive forces cause an abrupt shrinkage of the air gap and an exponential rise in tunneling current. This reduces the subthreshold swing further by a factor proportional to $dt_{air}/d\theta$, ultimately vanishing near the contact point, giving a corresponding low subthreshold swing. However removing the cantilever requires breaking the adhesive bonds at the drain, which requires a large negative bias and a corresponding large hysteresis that compromises the average swing. In a separate mechanism dominant for short relays such as ion channels, the presence of dipolar charges on the cantilever causes the subthreshold swing to be reduced by a factor $qt_{air}/\mu \propto t_{air}/L$, exploiting the property that many correlated charges are physically moving with the cantilever for the price of one.
\\\\
Looking at the overall transmission function, we see once again the role of gate tunability. Unlike TFETs where we are modulating the transmission bandwidth (section 3), or chiral TFETs where we modulate the transmission gap (section 4), in a NEMFET we modulate the height of the transmission function. As the cantilever is removed from the drain, its charges get depleted by the gate bias simultaneously as its tunnel transmission drops from unity to zero. 
\\\\
\\\\
{\subsection{6. Gating bandgap and effective mass: Mott switches?}} 
The physics of metal insulator transitions is complicated and varied among materials, involving both electronic transitions and structural phase transitions. An electronic Mott transition \cite{mott} has been postulated for materials such as VO$_2$. At its simplest, there is an increase in Coulomb forces relative to bandwidth with increased doping or decreasing pressure, leading to an abrupt opening of a gap in a metallic band and creation of a paramagnetic insulator. This correlation dominates for narrow d band materials. Depending on whether the repelled lower (Hubbard) d-band stays highest occupied, vs is shoved below an oxygen 2p-band, the material is labeled a Mott (e.g. VO$_2$) vs a charge transfer insulator (e.g. cuprates). 
\\\\
The tunability of the true bandgap (as opposed to transmission gap) leads to a sudden depletion of charge and a concomitant sharp drop in conductivity. The steepness of this curve is expected to show a low subthreshold swing \textcolor{black}{because of the appearance of a bandgap in the transmission function (Fig.~\ref{f5})}.
\textcolor{black}{Including this correlation effect quantitatively requires an approach beyond perturbative mean-field potentials to account for strong electron-electron repulsion \cite{9bhasko1,10bhasko2,agbook,agbook2}, but can be qualitatively captured with a model nonequilibrium electron response (Green's function), which then gives us the transmission in Landauer equation}
\\\\
\begin{figure}
\begin{center}
\includegraphics[width=3.3in,height=2.8in]{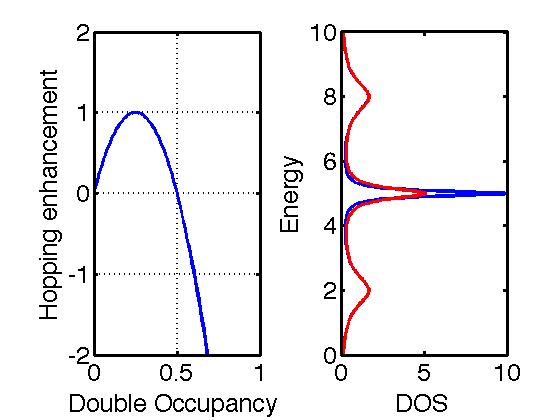}
\end{center}
\caption{\textcolor{black}{(Left) Renormalization of hopping with double occupancy leads to (b) a progressive vanishing of bandwidth of a resonant state at the Fermi energy (blue is with weak Coulomb, red with strong Coulomb), a divergence of effective mass with a continuous second order metal insulator transition. The transmission has the same general shape as the density of states, shown schematically above.}}
\label{br}
\end{figure}
Let us start with a single atom, modeled as a strongly correlated quantum dot. Such a dot shows pronounced Coulomb Blockade when we add a single charge of a given spin to it. The complementary spin state gets pushed up by Coulomb repulsion from the first spin, while the first spin state remains unchanged as it cannot feel a force due to itself. The self interaction correction can be included through an equation of motion technique that leads to the formation of local moments \cite{9bhasko1,10bhasko2,agbook,agbook2}, through a weighted probabilistic average between the filled charge and empty charge cases with and without repulsion. The one electron Green's function $G = (EI - H)^{-1}$ then amounts to \cite{agbook}
\begin{equation}
G^{LM}_\sigma(E) = \underbrace{\frac{1- n_{\displaystyle \bar{\sigma}}}{E-{\epsilon}_{\displaystyle {\sigma}}-\Sigma_{\displaystyle \sigma} }}_{\displaystyle\bar{\sigma}~empty} +
 \underbrace{\frac{ n_{\displaystyle \bar{\sigma}}}{E-{\epsilon}_{\displaystyle {\sigma}}-U-\Sigma_{\displaystyle \sigma}}}_{\displaystyle \bar{\sigma}~filled}
\label{LMG}
\end{equation}
where $\sigma$ is a particular spin (up or down) and $\bar{\sigma}$ is its compliment. The self-energy $\Sigma)\sigma$ accounts for open boundary imposed on Schr\"odinger equation at the contacts, and determines both the injection/removal rates and the level broadening $\gamma_\sigma$. 
The electron density is obtained self consistently with the density of states
\textcolor{black}{\begin{eqnarray}
n_{\displaystyle {\sigma}} &=&  \int D^{LM}_{\displaystyle {\sigma}}(E)f(E-E_F)dE \nonumber\\
D^{LM}_{\displaystyle {\sigma}} &=& -Im(G^{LM}_{\displaystyle {\sigma}})/\pi\nonumber\\
T^{LM}_{\displaystyle \sigma} &=& \gamma^{}_{\displaystyle \sigma}D^{LM}_{\displaystyle \sigma}/2,~~~\gamma^{}_{\displaystyle \sigma} = -2Im(\Sigma_{\displaystyle \sigma})
\label{LMscf}
\end{eqnarray}
}
As we deplete the electron density using a gate, the Coulomb potential is underscreened, with a Thomas-Fermi screening $U \sim U_0/(1 + k_s^2a_0^2)$ where $a_0$ is the Bohr radius, and $k_s$ is the screening wavevector $\sim n^{1/3}$ that reduces with reducing electron charge density. When the electron overlap becomes smaller than the Bohr radius, $n^{1/3}a_0 \ll  1  \rightarrow k_sa_0 \ll 1$, the unscreened Coulomb potential $U_0$ creates a gap that exceeds the bandwidth given by $\gamma^{}_{\displaystyle \sigma}$ and an electronically driven Mott metal insulator transition occurs. The kinetic energy for a free electron gas $3E_F/5 ~\sim n^{2/3}$ matches the Coulomb repulsion $q^2/r_s \sim n^{1/3}$ at that critical electron density, leading to the Mott criterion above. Fig.~\ref{f5} shows a simulation using the above equations, where we get a bandgap at a critical concentration.
\\\\
 The physics can be extended from a single quantum dot to a solid by treating the latter as an array of coupled quantum dots and replacing ${\epsilon}_{\displaystyle {\sigma}} \rightarrow {\epsilon}_{\displaystyle {k\sigma}}$. The Coulomb Blockade now morphs into a Mott insulator with two bands. The sudden removal of spectral weight at the contact Fermi energy leads to a sudden drop in conductivity, giving us a low subthreshold switch. 
\\\\
The modulation of bandgap increases the mobile charge density sharply and can be directly mapped onto the creation of a correlated electron liquid, expected to reduce the subthreshold swing analogous to the mechanism behind an ion channel relay, i.e., many charges moving for the price of one. {Experimentally, a purely electronic Mott transition is rarely seen in a transport measurement \cite{ramanathan1,ramanathan2}, and is usually complicated by concomitant thermally driven structural transitions. Part of the challenge is to realize good contacts and good electrostatic control. This is seen from the gate transfer factor $\alpha_G$ determining the effective field in the channel, and from the mode count $M = 2\pi\gamma D$, where $\gamma$ is the level broadening and $D$ is the density of states. The broadening follows Mathiessen's rule, $t = \hbar/\gamma = \hbar/\gamma_1 +\hbar/\gamma_2 + L/v$, so that the intrinsic channel velocity $v$ matters only when the contacts are excellent and their broadenings/injection rates are very high compared to the transit speed, $\gamma_{1,2} \gg \hbar v/L$}
\\\\
Photoemission experiments on $V_2O_3$ show an added feature over and above the opening of a Hubbard gap, namely, a secondary quasiparticle peak near the Fermi energy that keeps narrowing and losing spectral weight to the main Hubbard bands (Fig.~\ref{br}). The corresponding increase in effective mass m$^*$ from spectral narrowing slows down the electrons and provides an added mechanism for charge localization, known as the Brinkman-Rice transition. \textcolor{black}{In a simple 1D chain of atoms with strong Coulomb correlations, we can introduce a Gutzwiller projector that excludes the doubly occupied states from the uncorrelated many electron state $\Phi$ with an efficiency $g$, $\Psi_{\displaystyle {GW}} = g^D\Phi $ where $D =  \sum_i n_{\displaystyle i\uparrow}n_{\displaystyle i\downarrow}$ is the number of doubly occupied states. For a half filled chain, this projection renormalizes the hopping between atoms from $\gamma \rightarrow \gamma \times 8D(1-2D)$, plotted at the left of Fig.~\ref{br} and the Coulomb term to $UD$, and accordingly the transmission (Eq.~\ref{LMscf}). For an uncorrelated ground state $U = 0$ and $D = n_{\displaystyle i\uparrow} = n_{\displaystyle i\downarrow} = 1/4$ and the hopping is unaffected. However as we increase $U$ and thus $D > 1/4$, the hopping term reduces as $\gamma \propto (1 - U/U_c)^2$ till at $D = 1/2$
the hopping vanishes and the effective mass $m^*/m = 1/\gamma$ of the half filled state shoots up to infinity, creating an abrupt first order phase transition \cite{agbook},~\cite{11gebhard},~\cite{fulde}. Note that unlike the Hubbard gap where there are no states at the Fermi energy, in the Brinkman-Rice transition the state at the Fermi energy stays half filled but becomes infinitely narrow and thus ultimately insulating \cite{12dmft}.}
\\\\
{\subsection{Summary}} 
The aim of this paper was to provide simplified models that explain how a gate dependent transmission function leads to an added contribution to current flow, lowering the switching voltage and ultimately the subthreshold swing. \textcolor{black}{Such gate dependent transmissions take varied forms, from bandwidth modulation (Brinkman-Rice, Fig.~\ref{br}) to bandgap modulation (Mott-Hubbard, Eq.~\ref{LMscf}, Fig.~\ref{f5}) to transport gap modulation (chiral flow across GPNJs, Eqs.~\ref{Tpnj},~\ref{tc}, Figs.~\ref{f2},~\ref{f3}) to amplitude modulation (NEMFETs, Eq.~\ref{tnems}, Fig.~\ref{f4}) to both bandwidth and amplitude modulation (tunnel FETs, Eq.~\ref{twkb},~Fig.~\ref{f1})}. Analogous arguments can probably be made for devices not covered in this writeup, like polaron mediated switches driven by dimerization (\textcolor{black}{a transmission shift away from the Fermi window by the polaron formation energy}), or negative capacitance devices where a second order ferroelectric phase transition causes the capacitance curve $Q-V_g$  to flip the sign of its curvature and make the gate transfer factor $\alpha_G > 1$ \cite{14sayeef}, \textcolor{black}{giving a voltage upconversion and a transmission shift larger than the applied bias}.  
{\subsection{Acknowledgements}} 
AWG acknowledges funding support from the Nano Research Initiative (NRI).

\end{document}